\documentclass[final,authoryear,3p,times,twocolumn]{elsarticle}

\usepackage[latin2]{inputenc}
\usepackage{graphicx}

\usepackage{amssymb}
\usepackage{amsthm}
\usepackage{ulem}

\journal{Applied Radiation and Isotopes 110(2016)109}

\begin{document}

\begin{frontmatter}



\title{Excitation functions for(d,x)reactions on $^{133}$Cs up to $E_d = 40$ MeV}


\author[1]{F. T\'ark\'anyi}
\author[1]{F. Ditr\'oi\corref{*}}
\author[1]{S. Tak\'acs}
\author[2]{A. Hermanne}
\author[3]{M. Baba}
\author[4]{A.V. Ignatyuk}

\cortext[*]{Corresponding author: ditroi@atomki.hu}

\address[1]{Institute for Nuclear Research, Hungarian Academy of Sciences (ATOMKI),  Debrecen, Hungary}
\address[2]{Cyclotron Laboratory, Vrije Universiteit Brussel (VUB), Brussels, Belgium}
\address[4]{Institute of Physics and Power Engineering (IPPE), Obninsk, Russia}
\address[3]{Cyclotron Radioisotope Center (CYRIC), Tohoku University, Sendai, Japan}

\begin{abstract}
In the frame of a systematic study of excitation functions of deuteron induced reactions  the excitation functions of the $^{133}$Cs(d,x)$^{133m,133mg,131mg}$Ba, ${134,132}$Cs and $^{129m}$Xe nuclear reactions were measured up to 40 MeV deuteron energies by using the stacked foil irradiation technique and $\gamma$-ray spectroscopy of activated samples. The results were compared with calculations performed with the theoretical nuclear reaction codes ALICE-IPPE-D, EMPIRE II-D and TALYS calculation listed in the TENDL-2014 library.  A moderate agreement was obtained. Based on the integral yields deduced from our measured cross sections, production of $^{131}$Cs via the $^{133}$Cs(d,4n)$^{131}$Ba $\longrightarrow$ $^{131}$Cs reaction and $^{133}$Ba via $^{133}$Cs(d,2n) reactions is discussed in comparison with other charged particle production routes.
\end{abstract}


\end{frontmatter}


\section{Introduction}
\label{1}
Cross sections of charged particle induced reactions in Cs play an important role for production of radioisotopes of Ba and Cs, used in different applications. The nuclear data are needed for production of long-lived $^{133}$Ba, used as standard source for calibration of gamma spectrometers and for many other applications \citep{Meulen, Walt}. Shorter-lived $^{134m}$Cs was used for myocardial imaging and for investigation of transport and metabolism process of cesium and as a substitute for potassium \citep{Eybel}. The pure EC decaying $^{131}$Cs (X-ray emitter) gained application in internal radio-therapy \citep{Henschke, Murphy} . 
We have started earlier a systematic investigation on the possibility of production $^{131}$Cs at a cyclotron. The study includes series of new experimental measurements of proton, deuteron and alpha induced reactions on $^{nat}$Cs, $^{nat}$Ba, $^{131}$Xe and $^{nat}$Xe, and a critical evaluation of the experimental data reported in the literature \citep{TF2010, TF2009b, TF2008b, TF2009, TF2008}.  Simultaneously with the investigation of the production routes of medical isotopes a systematical study of the activation cross section data of deuteron induced nuclear reactions is also in progress including most of the elements, to produce an activation data library \citep{Hermanne,  TF2011, TF2007}. 
A literature search for activation data of deuteron induced reactions on Cs resulted in identification of some publications on experimental cross sections and yield data.
Baron reported cross section for production of  $^{134m}$Cs by 19.4 MeV deuterons \citep{Baron}. Pement measured  cross sections for the $^{133}$Cs(d,2n) reaction in the  4-15 MeV energy range \citep{Pement}. Mocoroa studied isomeric ratios for the $^{133}$Cs(d,2n) and $^{133}$Cs(d,p) reactions in  the 9.4-28.5 MeV energy range \citep{Mocoroa}. Natowitz measured $^{134m}$Cs/$^{134g}$Cs isomeric and yield ratios in the 2.5-19.4 MeV deuteron energy range \citep{Natowitz} . Dmitriev published cross section ratios for $^{133}$Ba (9.1-22.4 MeV) \citep{Dmitriev73}. Neirincky measured integral yield at 16.7 MeV incident energy for production of $^{133m}$Ba \citep{Neirincky}. Dmitriev, in systematic investigations of production yields at 22 MeV deuteron energy for cesium targets, measured the integral yields for $^{133}$Ba,$^{133m}$Ba,$^{123}$Xe and $^{134}$Cs \citep{Dmitriev83}. Mukhammedov measured integral yield data for production of $^{133m}$Ba, $^{134}$Cs as a function of incident energies from 7 to 12 MeV deuterons \citep{Mukhammedov}.

\section{Experimental}
\label{2}

The excitation functions were measured up to 40 MeV via the activation technique by bombarding stacked CsCl targets with a low intensity deuteron beam at the AVF-930 cyclotron of the Tohoku University (Sendai). Reactions induced on Al monitor foils were used to refine the parameters (energy, intensity) of the incident beam. 
Special care was taken in preparation of uniform targets with well-known thickness, in determination of the energy and intensity of the bombarding beam along the target stack and in determination of the activities of the samples. Self-supporting pellets (120 $mg/cm^2$) of CsCl targets were made by pressing. The pellets were then stacked with 100 $\mu$m thick Al backings and 10 $\mu$m thick Al protecting covers. 
To minimize the water content of the targets, the stock CsCl powder (99.9 \%) was dried under vacuum and the pellets were sealed in a plastic bag before irradiation. The targets were irradiated in a He gas atmosphere. For determination of the beam intensity and energy, the complete excitation function of the monitor reactions were re-measured simultaneously by using the cover (10 $\mu$m) and the 100 $\mu$m backing Al-foils. The target stack was irradiated with 40 MeV incident energy for about a 40 min, at 25 nA beam current. The target integrity after the irradiation was checked. Final flux data were determined from the monitor reactions. The energy of the extracted beam was determined by a calibrated magnetic bending set-up. Possible effect of secondary neutrons was experimentally checked by inserting an additional pellet beyond the charged particle stopping range. 
The activity of the irradiated samples was measured without chemical separation by using well-proved gamma-spectroscopy. Due to the high initial dose rate the first activity measurements were started around one day after the end of bombardment, which resulted in the complete decay of most of the short-lived radioisotopes. The second series of gamma spectra measurement started   four days after EOB. 
The energy degradation along the stack was determined by calculation and by comparison with the re-measured $^{27}$Al(d,x)$^{24}$Na  monitor reactions \citep{Andersen, TF1991, TF2001} (Fig. 1).
 The uncertainty of the energy scale was estimated by taking into account the energy uncertainty of the primary beam, the possible variation in the target thickness and the effect of beam straggling. 
The cross sections were deduced by using the standard activation formula, so called isotopic cross sections were calculated taking into account that Cs is monoisotopic. The decay data, the contributing reactions and the reaction Q-values for production of the measured radioisotopes of Ba, Cs, Xe and their longer-lived isomeric states, taken from \citep{Kinsey, Pritychenko, Sonzogni}, are summarized in Table 1.
The uncertainties of the cross-sections were estimated by using the error propagation formula used for calculation \citep{Error}. The uncertainties of the contributing factors were: number of bombarding particles (7 \%), gamma intensity data (1-3 \%), detector efficiency (5 \%), peak area (0.1 to 10 \%), number of the target nuclei (5 \%). Typical overall uncertainties of the cross-sections are around 12-15 \%.
 
\begin{figure}
\includegraphics[width=0.5\textwidth]{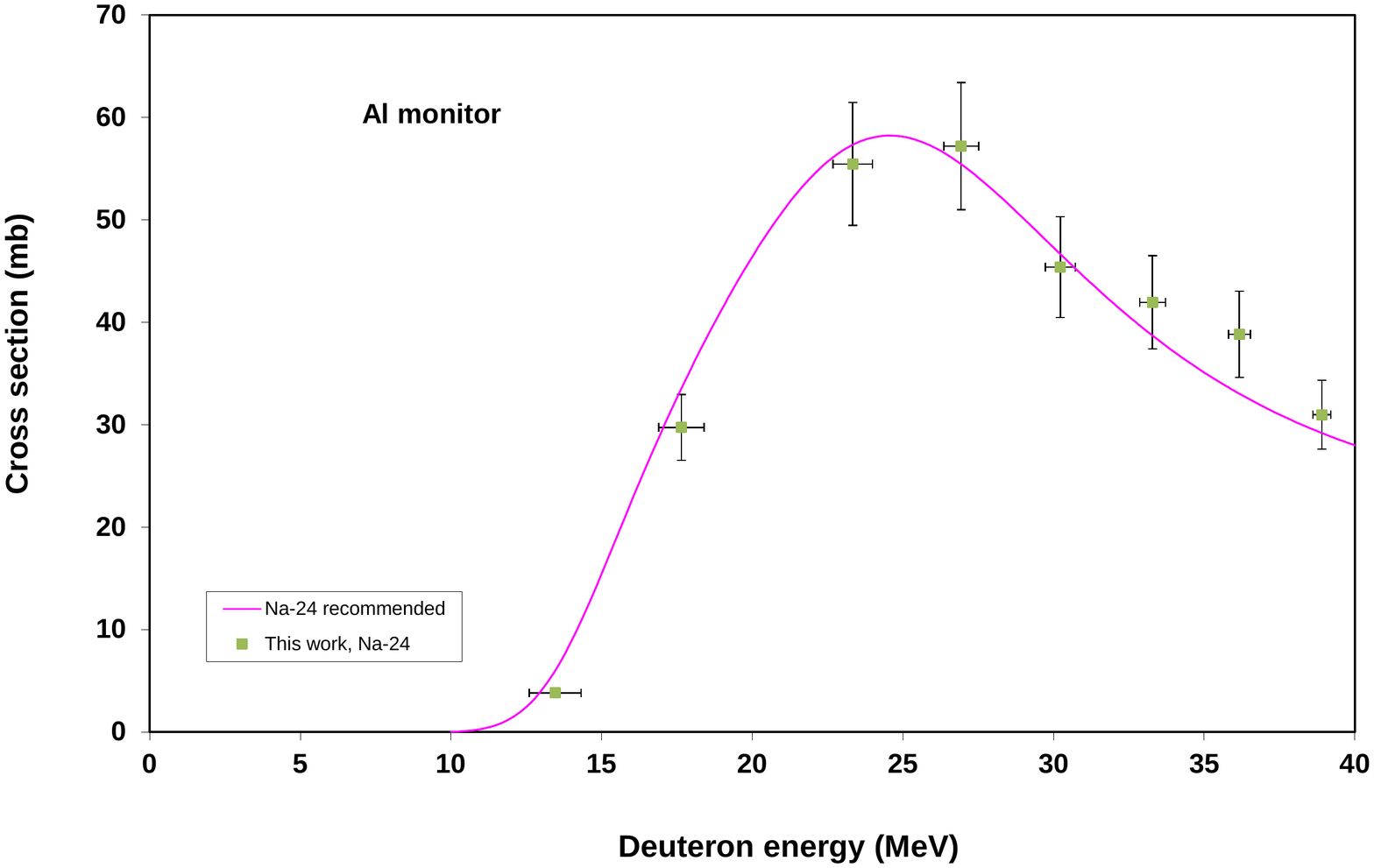}
\caption{Monitoring of the deuteron beam parameters with the $^{nat}$Al(d,x)$^{24}$Na reactions and comparison with the recommended values}
\label{fig:1}       
\end{figure}

\begin{table*}[t]
\tiny
\caption{Decay characteristic of the investigated reaction products $^{133m,133mg,131mg}$Ba, $^{134,132}$Cs, and $^{129m}$Xe}
\begin{center}
\begin{tabular}{|p{0.5in}|p{0.6in}|p{0.6in}|p{0.5in}|p{0.6in}|p{0.8in}|p{0.6in}|} \hline 
\textbf{Nuclide\newline J$^{\pi}$\newline level(keV)} & \textbf{Half-life\newline } & \textbf{Decay\newline mode (\%)} & \textbf{E$_{\gamma}$ (keV)} & \textbf{I$_{\gamma}$ (\%)} & \textbf{Contributing\newline process} & \textbf{Q-value\newline (keV)} \\ \hline 
${}^{133m}$Ba\newline 11/2${}^{-}$\newline 288.252 & 38.93 h & IT:99.9896~\newline EC:0.0104 & 275.925\newline \newline  & 17.69\newline \newline  & (d,2n) & -3524.266\newline  \\ \hline 
${}^{133}$Ba\newline 1/2${}^{+}$ & 10.551 a\newline \newline  & EC:100 & 276.3989\newline 302.8508\newline 356.0129\newline 383.8485 & 7.16\newline 18.34\newline 62.05\newline ~8.94 & (d,2n) & -3524.266\newline \newline \newline  \\ \hline 
${}^{131m}$Ba\newline 9/2${}^{-}$ & 14.6 min\newline  & IT:100 & 108.45\newline  & 55.44\newline  & (d,4n) & -20536.57\newline  \\ \hline 
${}^{131}$Ba\newline $\frac{1}{2}^{+}$ & 11.5 d\newline  & EC:100 & 123.804\newline 216.088\newline 373.256\newline 496.321\newline  & ~29.8\newline 20.4\newline 14.40\newline 48.0\newline  & (d,4n) &  -20536.57\newline  \\ \hline 
${}^{134m}$Cs\newline 8$^{-}$\newline 138.7441 & 2.912 h & IT:100 & 127.502 & 12.6 & (d,n) & 5943.524 \\ \hline 
${}^{134}$Cs\newline 4$^+$ & 2.0652 a & $\betaup$${}^{-}$:99.9997\newline $\varepsilonup$: 0.0003 & 563.246\newline 569.331\newline 604.721\newline 795.864\newline 801.953 & 8.338\newline 15.373\newline 97.62\newline 85.46\newline 8.688 & (d,n) & 5943.524 \\ \hline 
${}^{132}$Cs\newline 2${}^{+}$ & 6.479 d\newline  & $\betaup$${}^{-}$: 1.87\newline EC: 98.13 & 667.714\newline  & 97.59\newline  & (d,p2n), & -11210.57\newline  \\ \hline 
${}^{131}$Cs\newline 5/2${}^{+}$ & 9.689 d\newline  &  EC:100 &  &  & (d,p3n)\newline ${}^{131}$Ba decay & -18379.26\newline \newline  \\ \hline 
${}^{127}$Cs\newline 1/2+\newline  & 6.25 h & $\betaup$${}^{+}$:100 & ~124.70\newline 411.95\newline 462.31\newline 587.01 & 11.38\newline 62.9\newline ~5.08\newline 4.21 & (d,p7n)\newline  & -52483.27 \\ \hline 
${}^{129m}$Xe\newline 11/2${}^{-}$\newline 236.14 & 8.88 d & IT:100 & ~196.56 & ~4.59 & (d,2p4n) & -33102.375 \\ \hline 
\end{tabular}

\end{center}
\begin{flushleft}
\tiny{\noindent Increase the Q-values if compound particles are emitted by: np-d, +2.2 MeV; 2np-t, +8.48 MeV; n2p-$^3$He, +7.72 MeV; 2n2p-$\alpha$, +28.30 MeV. In the case of excited states decrease the Q-value with the level energy (higher threshold).
}
\end{flushleft}

\end{table*}




\section{Theoretical calculations}
\label{3}
The cross sections of the investigated reactions were calculated using the pre-compound model codes ALICE-IPPE \citep{Dityuk} and EMPIRE-II \citep{Herman}  modified for deuterons by Ignatyuk (D versions) \citep{Ignatyuk}. The theoretical curves were determined using one recommended input data-set \citep{Belgya} without any optimization or adjustment of parameters to the individual reactions or stable target isotopes. Independent data for isomers with ALICE-D code was obtained by using the isomeric ratios calculated with EMPIRE.
The experimental data are also compared with the cross section data reported in the TENDL-2014 \citep{Koning2014} nuclear reaction data library. The TENDL library was developed on the base of the TALYS nuclear model code system \citep{Koning2012} for direct use in both basic physics and applications (default TALYS calculations).

\section{Experimental results}
\label{4}
The numerical values of the recently measured cross-sections of the investigated reactions are collected in Table 2. The excitation functions are shown in Figs. 2-8. 

\subsection{The $^{133}$Cs(d,2n)$^{133m}$Ba, $^{133g}$Ba(m+) reactions}
\label{4.1}
No cross section data have been earlier reported for production of the $^{133m}$Ba (T$_{1/2}$ = 38.9 h) metastable state (Fig. 2). The values in TENDL-2014, the calculated results of ALICE-D and EMPIRE-D all overestimate the experimental data in different proportions.
We could find only one earlier experimental data set for the cumulative activation cross section of the ground state of $^{133g}$Ba (T$_{1/2}$= 10.52 a) after the decay of the isomeric state, reported by \citep{Pement}. There is an acceptable agreement with our results in the overlapping energy region (Fig. 3). There are large differences in magnitude between the TENDL-2014 theoretical results and the experimental data. The predictions of EMPIRE-D are rather well describing the experimental values over the whole studied energy range, while the ALICE-D values are decreasing too fast above 12 MeV.

\begin{figure}
\includegraphics[width=0.5\textwidth]{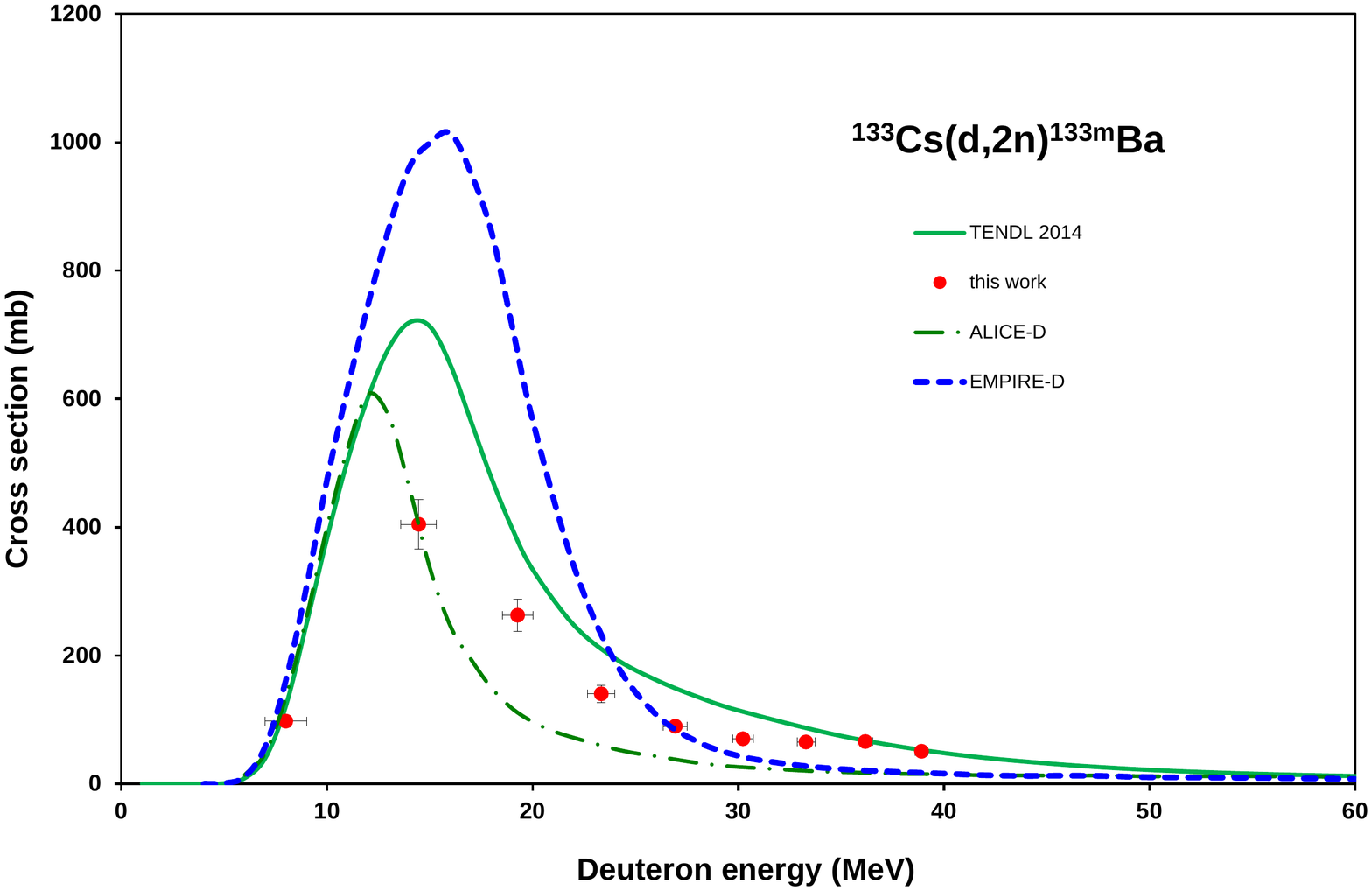}
\caption{Excitation function of the $^{133}$Cs(d,2n)$^{133m}$Ba reaction}
\label{fig:2}       
\end{figure}

\begin{figure}
\includegraphics[width=0.5\textwidth]{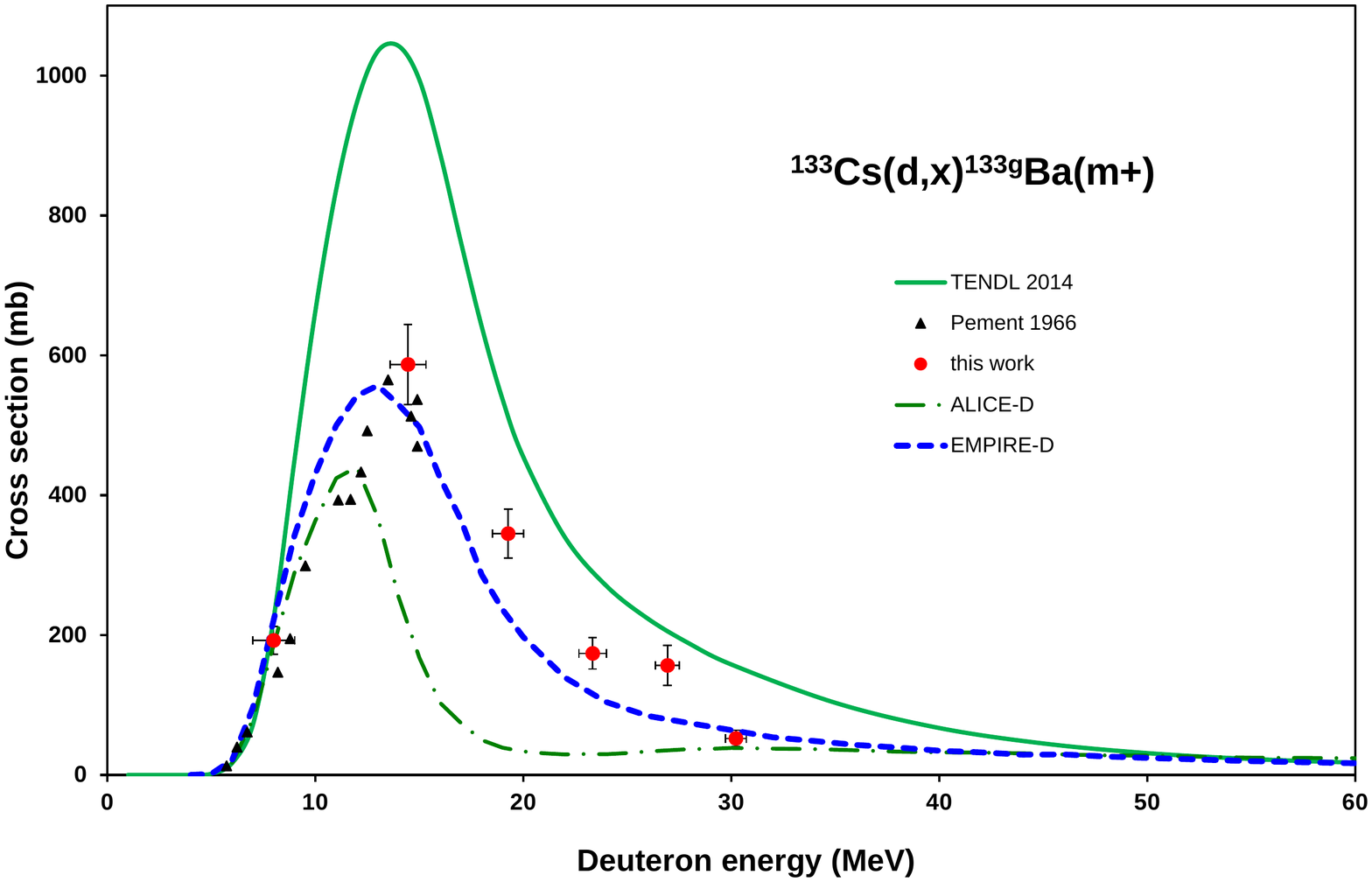}
\caption{Excitation function of $^{133}$Cs(d,2n)$^{133g}$Ba(m+) reaction}
\label{fig:3}       
\end{figure}

\subsection{The $^{133}$Cs(d,4n)$^{131g}$Ba(m+) reaction}
\label{4.2}
Due to the long cooling time we could not assess cross section data for the short-lived $^{131m}$Ba metastable state (T$_{1/2}$ = 14.6 min). 
The cross sections for $^{131g}$Ba (T$_{1/2}$ = 11.5 d) obtained after total decay of the isomeric state (m+) are shown in Fig. 4. The model calculation results in TENDL-2014 support our result both in shape and in magnitude. The EMPIRE-D and ALICE-D are overestimating the experimental values and show opposite energy shifts of the maximum position.

\begin{figure}
\includegraphics[width=0.5\textwidth]{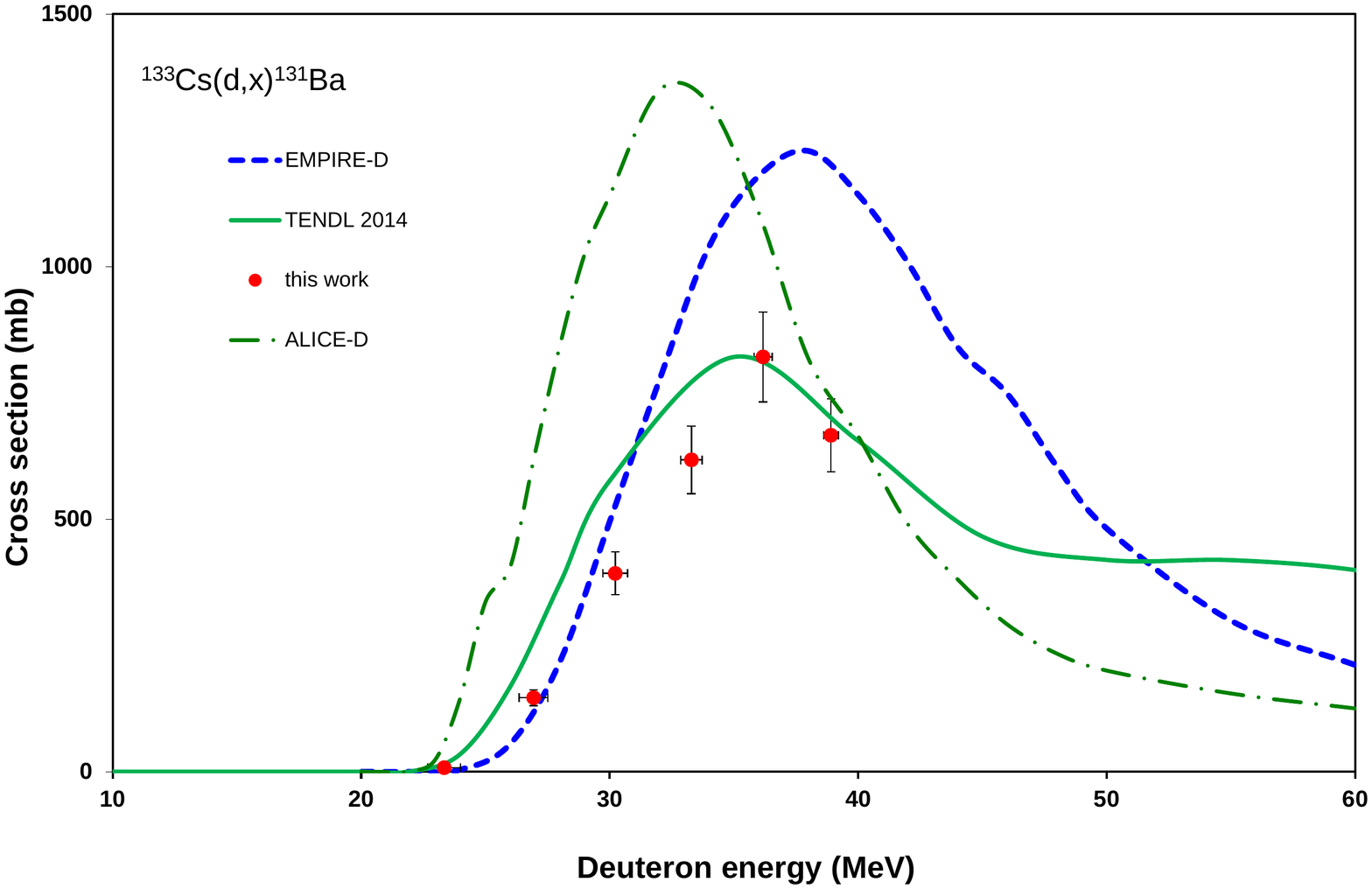}
\caption{Excitation function of $^{133}$Cs(d,4n)$^{131g}$Ba(m+) reaction}
\label{fig:4}       
\end{figure}

\subsection{The $^{133}$Cs(d,p)$^{134g}$Cs(m+) reaction}
\label{4.3}
The excitation function for production of $^{134g}$Cs (Fig. 5) was measured after the total decay of the short-lived meta-stable state (T$_{1/2}$ = 2.912 h, IT = 100\%) to the long-lived ground state (T$_{1/2}$ = 2.0652 a). The agreement with the rather scattered data of \citep{Natowitz} is reasonable, where it is possible to compare. The deep valley around 10 MeV in Natowitz results could be reproduce neither with our measurement (no data in this region) nor with the model codes. The significant underestimation of the experimental (d,p) cross section data in the TENDL-2014 library can be observed also in this case. The values predicted by ALICE-D and EMPIRE-D are overestimating the maximum by a factor of two. In Fig. 5 we also present the predictions for (d,p) reactions coming from systematics in this mass region.

\begin{figure}
\includegraphics[width=0.5\textwidth]{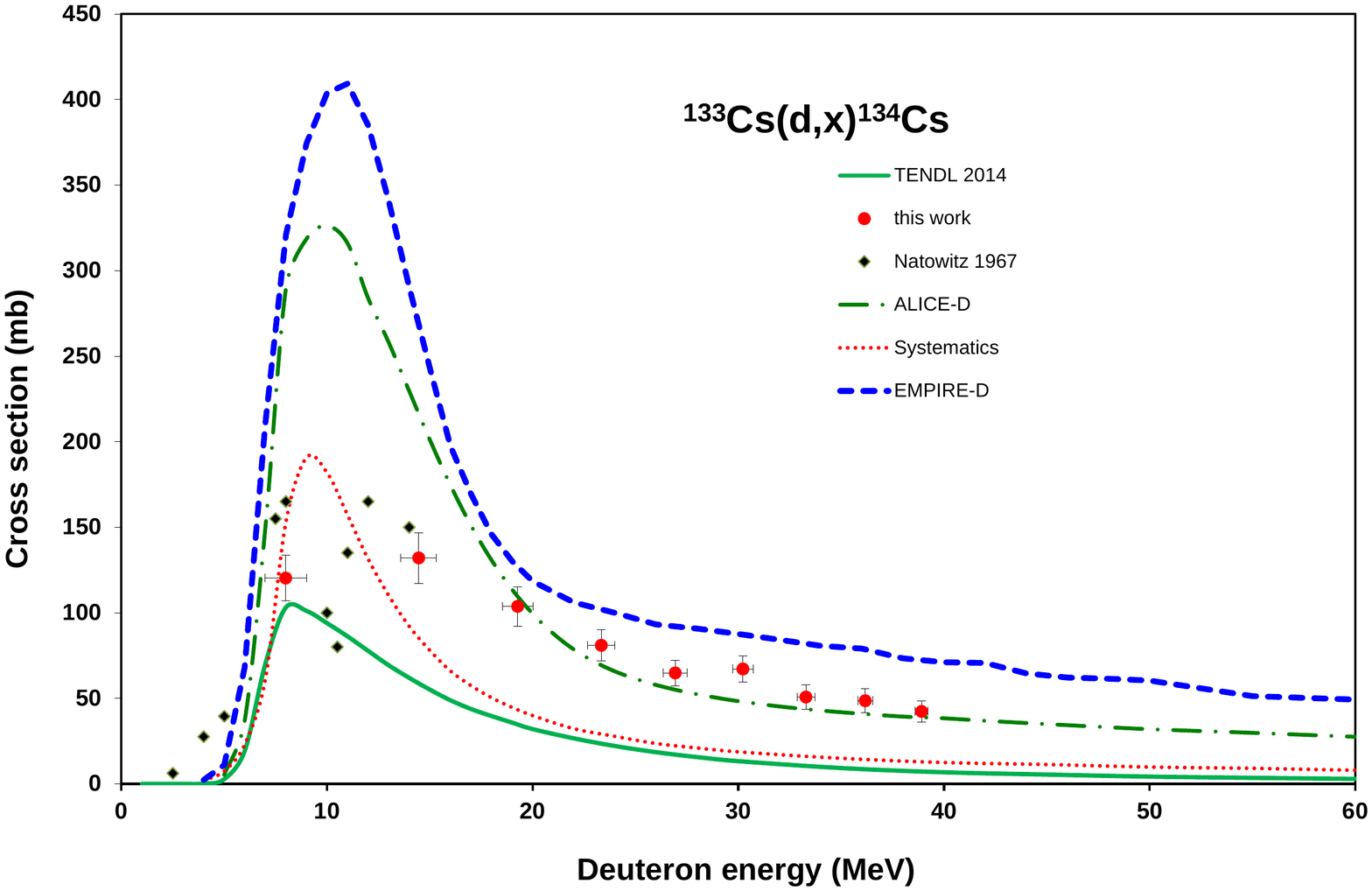}
\caption{Excitation function of $^{133}$Cs(d,p)$^{134g}$Cs(m+) reaction}
\label{fig:5}       
\end{figure}

\subsection{The $^{133}$Cs(d,p2n)$^{132}$Cs reaction}
\label{4.4}
The measured cross section data of $^{132}$Cs (T$_{1/2}$ = 6.479 d) are shown in Fig. 6 in comparison with the theoretical predictions of TENDL-2014, ALICE-D and EMPIRE-D. While agreement for the TALYS based result is acceptable, the two other codes strongly underestimate the experimental data. No earlier cross section data are available. 

\begin{figure}
\includegraphics[width=0.5\textwidth]{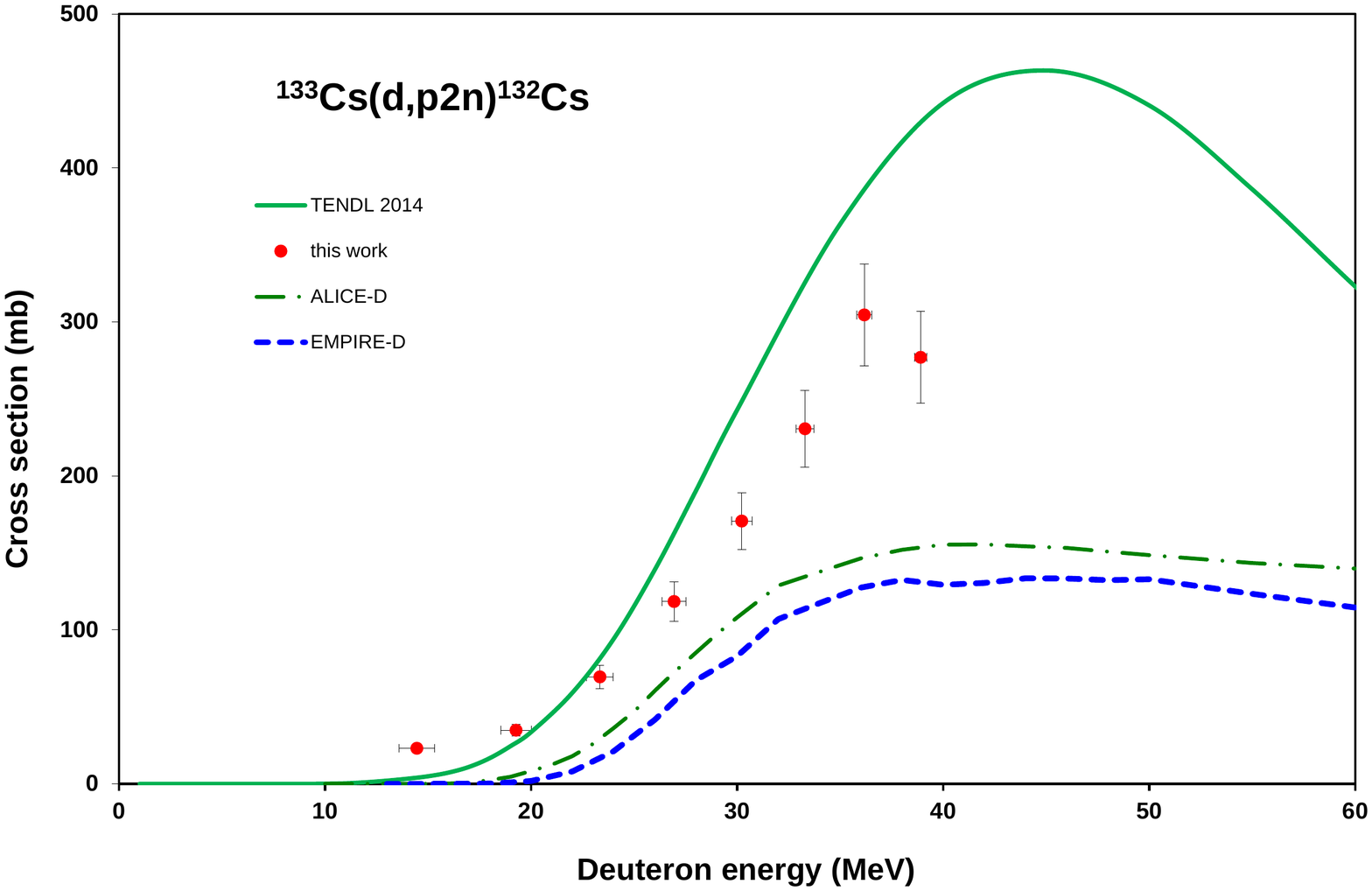}
\caption{Excitation function of $^{133}$Cs(d,p2n)$^{132}$Cs reaction}
\label{fig:6}       
\end{figure}

\subsection{The $^{133}$Cs(d,p3n)$^{131}$Cs reaction}
\label{4.5}
We could not measure the direct production of $^{131}$Cs (T$_{1/2}$ = 9.689 d) as this radionuclide decays without emission of gammas and can only be quantified by a dedicated series of measurements of the complex X-ray spectrum. To estimate the possibilities for direct formation of the medically important $^{131}$Cs we present in Fig. 7 the predictions of our ALICE-D and EMPIRE-D calculations together with the TALYS results in the TENDL-2014 library. 

\begin{figure}
\includegraphics[width=0.5\textwidth]{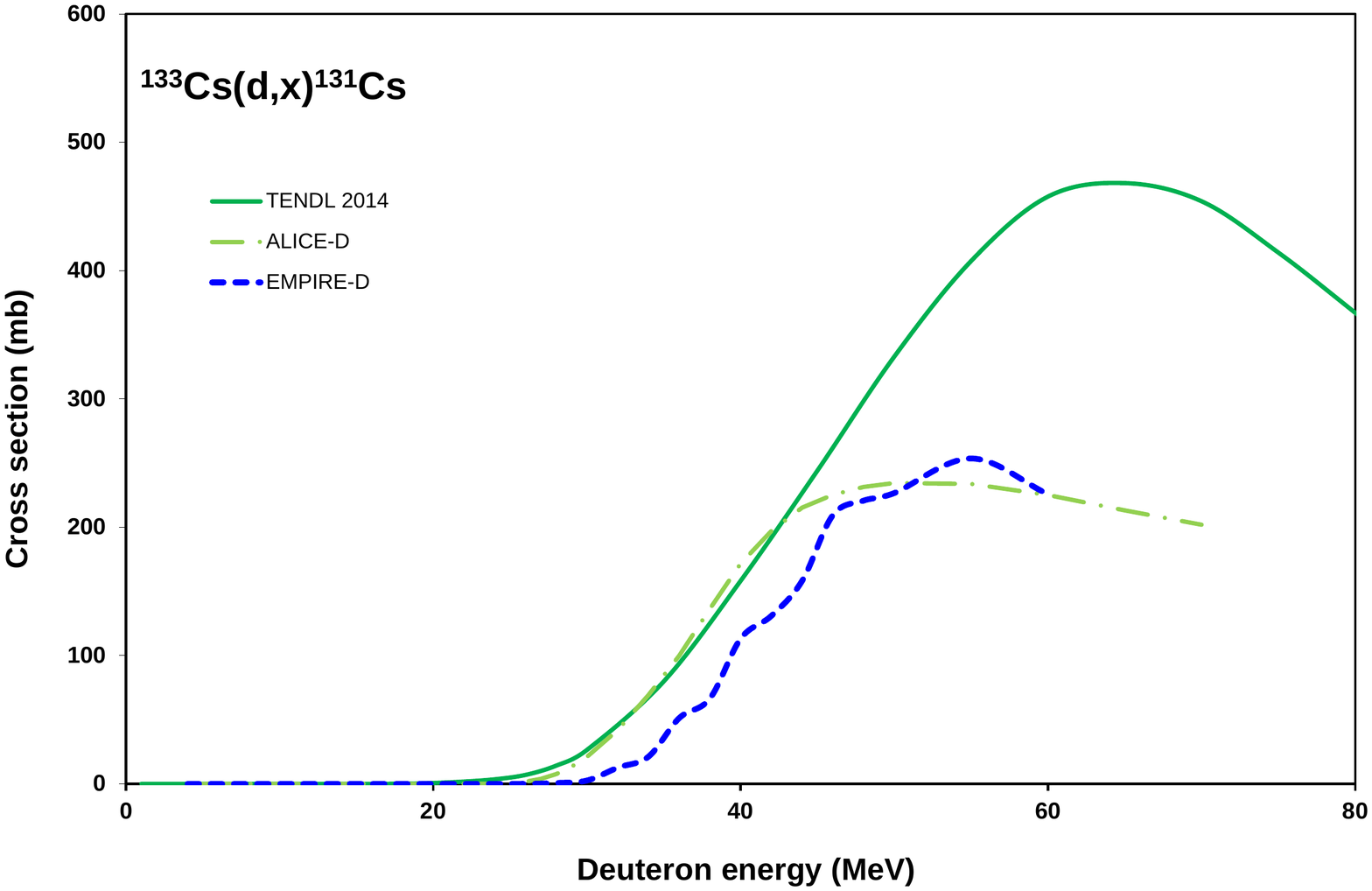}
\caption{Theoretical excitation functions of $^{133}$Cs(d,p3n)$^{131}$Cs reaction}
\label{fig:7}       
\end{figure}

\subsection{The $^{133}$Cs(d,2p6n)$^{129m}$Xe}
\label{4.6}
The radio isotope $^{129}$Xe has two longer-lived isomeric states:  the T$_{1/2}$ = 8.88 d half-life metastable state and the stable ground state. The practical threshold of 20 MeV shows that $^{129m}$Xe is produced at lower energy directly via a $^{133}$Cs(d,$\alpha$2n) reaction and not by emission of independent nucleons. The measured excitation function for production of $^{129m}$Xe is shown in Fig. 8 in comparison with the theoretical results. Here the results of ALICE-D and TENDL-2014 describe well the shape and values of our few experimental points. EMPIRE-D seems to be too large, but it reflects the effect of alpha-channels, description, which differs very strongly in ALICE and EMPIRE (different pre-equilibrium models).

\begin{figure}
\includegraphics[width=0.5\textwidth]{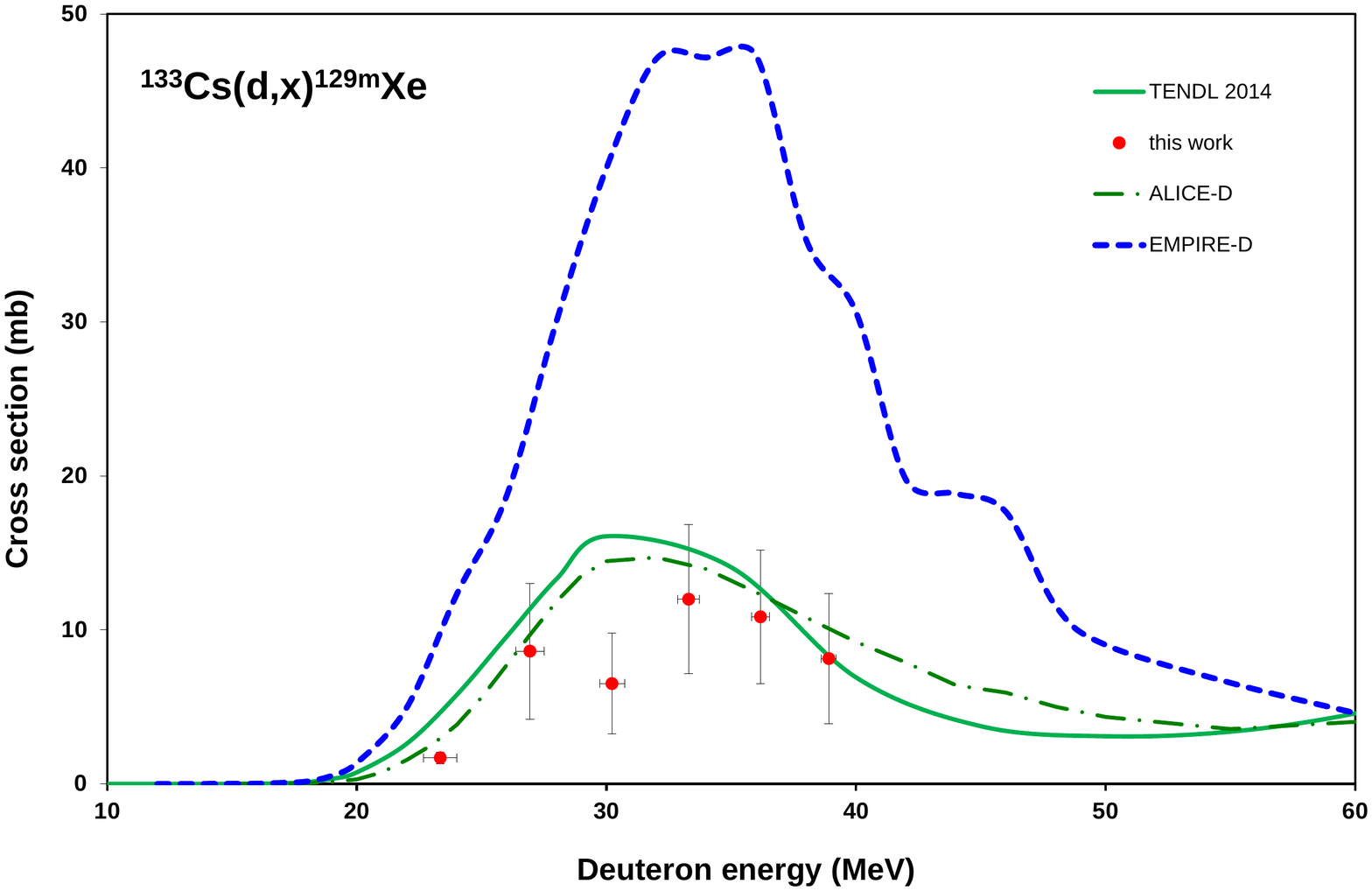}
\caption{Excitation function of the $^{133}$Cs(d,2p6n)$^{129m}$Xe reaction}
\label{fig:8}       
\end{figure}

\begin{table*}[t]
\tiny
\caption{Experimental cross sections of the investigated reaction products:}
\begin{center}
\begin{tabular}{|p{0.2in}|p{0.2in}|p{0.3in}|p{0.2in}|p{0.3in}|p{0.2in}|p{0.3in}|p{0.2in}|p{0.3in}|p{0.2in}|p{0.3in}|p{0.2in}|p{0.3in}|p{0.2in}|} \hline 
\multicolumn{2}{|c|}{\textbf{Deuteron energy}} & \multicolumn{2}{|c|}{\textbf{${}^{131g}$Ba}} & \multicolumn{2}{|c|}{\textbf{${}^{133m}$Ba}}  & \multicolumn{2}{|c|}{\textbf{${}^{133g}$Ba}} & \multicolumn{2}{|c|}{\textbf{${}^{132g}$Cs}} & \multicolumn{2}{|c|}{\textbf{${}^{134g}$Cs}}  & \multicolumn{2}{|c|}{\textbf{${}^{129m}$Xe}}  \\ \hline 
\multicolumn{2}{|c|}{\textbf{MeV}} & \multicolumn{12}{|c|}{\textbf{mb}} \\ \hline 
\textbf{E} & \textbf{dE} & \multicolumn{12}{|c|}{\textbf{$\sigma \pm d\sigma$}} \\ \hline 
38.9 & 0.3 & 666.1 & 72.1 & 50.9 & 4.9 & ~ &  & 277.1 & 30.0 & 42.4 & 6.1 & 8.1 & 4.2 \\ \hline 
36.2 & 0.4 & 821.1 & 88.8 & 65.9 & 6.3 & ~ &  & 304.5 & 32.9 & 48.6 & 7.1 & 10.9 & 4.3 \\ \hline 
33.3 & 0.4 & 617.3 & 66.8 & 65.3 & 6.2 & ~ &  & 230.6 & 24.9 & 50.7 & 7.3 & 12.0 & 4.8 \\ \hline 
30.2 & 0.5 & 392.8 & 42.5 & 70.3 & 6.7 & 52.2 & 11.9 & 170.7 & 18.5 & 67.1 & 7.7 & 6.5 & 3.3 \\ \hline 
26.9 & 0.6 & 146.5 & 15.9 & 89.8 & 8.5 & 156.6 & 28.8 & 118.4 & 12.8 & 64.8 & 7.5 & 8.6 & 4.4 \\ \hline 
23.3 & 0.7 & 8.1 & 0.9 & 140.2 & 13.3 & 173.8 & 22.4 & 69.4 & 7.5 & 81.0 & 9.2 & 1.7 & 0.4 \\ \hline 
19.3 & 0.8 & ~ &  & 262.8 & 24.9 & 345.2 & 35.1 & 34.8 & 3.8 & 103.8 & 11.5 & ~ &  \\ \hline 
14.5 & 0.9 & ~ &  & 404.5 & 38.4 & 586.8 & 57.4 & 23.1 & 2.5 & 132.0 & 14.8 & ~ &  \\ \hline 
8.0 & 1.0 & ~ &  & 97.6 & 9.3 & 192.4 & 19.9 & 5.4 & 0.6 & 120.3 & 13.3 & ~ &  \\ \hline 
0.6 & 1.2 & ~ &  & 0.1 & 0.01 & ~ &  & 3.7 & 0.4 & 1.0 & 0.1 & ~ &  \\ \hline 
\end{tabular}

\end{center}
\end{table*}

\section{Integral yields}
\label{5}
Based on the experimentally determined cross sections we have calculated the differential and integral yields of the produced radio-isotopes. The yields are so called physical yields, calculated for an instantaneous irradiation \citep{Bonardi, Otuka}. 
Figs. 9 and 10 show the integral yields for the production of the investigated radionuclides of Ba, Cs and Xe in comparison with the few experimental integral yield data in the literature (see literature review in the introduction). The experimental thick target yield are in good agreement in the case of $^{133m}$Ba and $^{134g}$Cs with our yield data deduced from experimental cross sections.  In the case of $^{133g}$Ba the previous data are somewhat higher, and there are no previous measurements for the rest of the radioisotopes.

\begin{figure}
\includegraphics[width=0.5\textwidth]{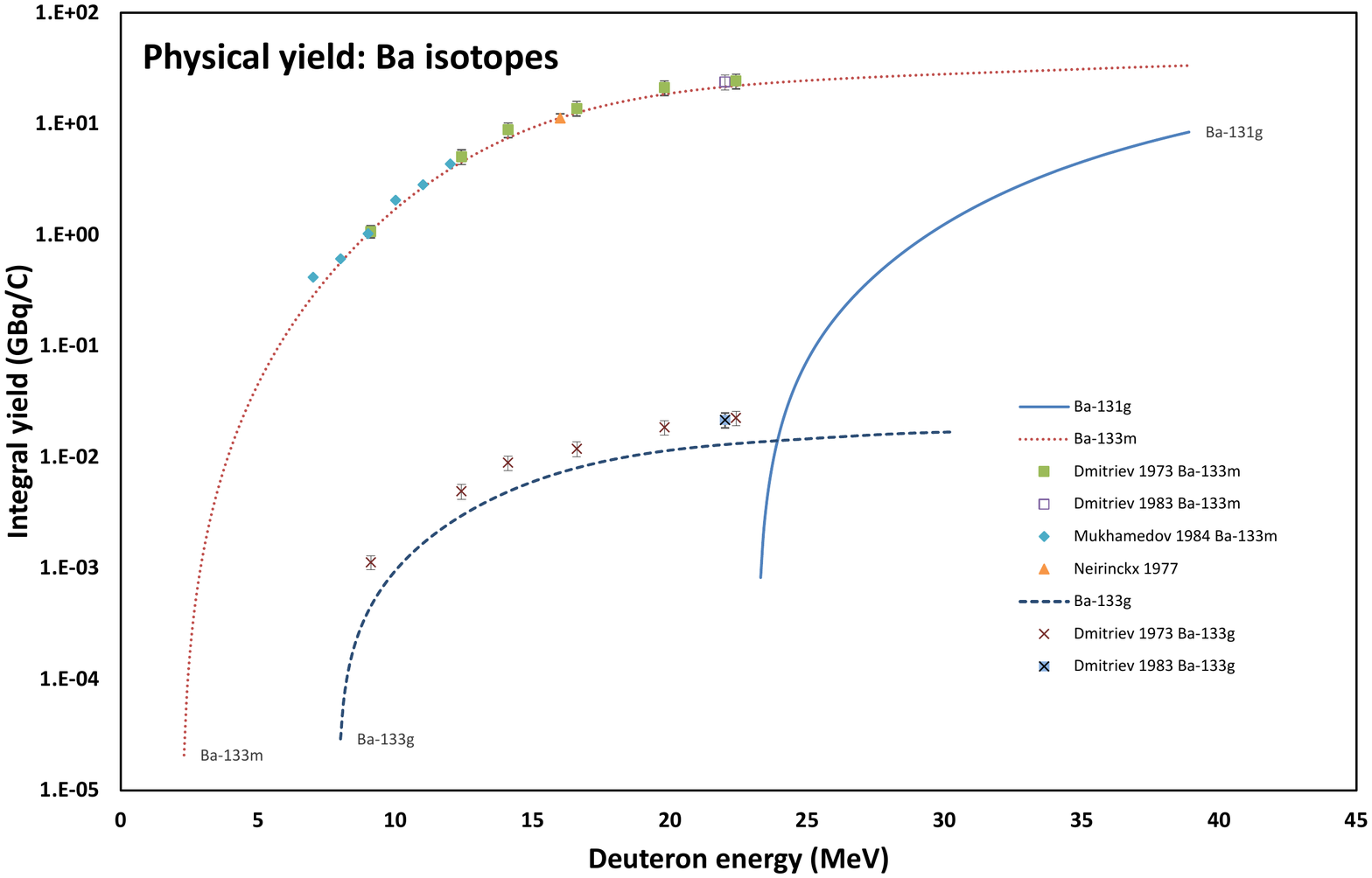}
\caption{Integral yields for production of radioisotopes of Ba by bombarding $^{133}$Cs  with deuteron}
\label{fig:9}       
\end{figure}

\begin{figure}
\includegraphics[width=0.5\textwidth]{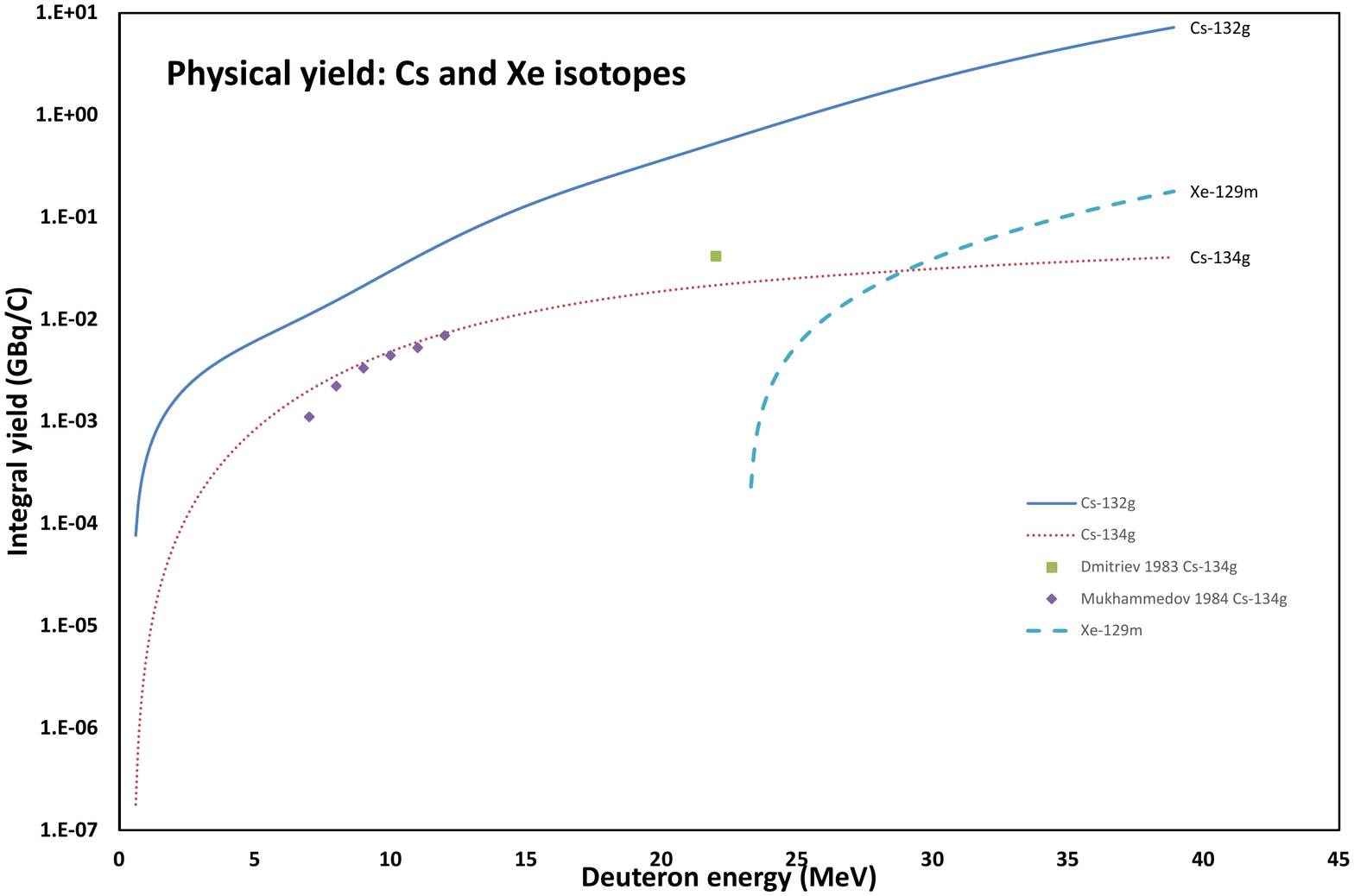}
\caption{Integral yields for production of radioisotopes of Cs and Xe by bombarding $^{133}$Cs  with deuterons}
\label{fig:10}       
\end{figure}

\section{Production of $^{131}$Cs and $^{133}$Ba}
\label{6}

\subsection{$^{131}$Cs}
\label{6.1}
As it was discussed in more detail in our previous paper \citep{TF2009b} for commercial use $^{131}$Cs is presently produced at high flux nuclear reactors by radioactive decay of its $^{131}$Ba mother obtained through neutron capture on naturally occurring $^{130}$Ba (natural Ba contains only 0.106 \% $^{130}$Ba) or on enriched $^{130}$Ba targets. As it was mentioned, we have studied already a few charged particle induced production routes for $^{131}$Ba/$^{131}$Cs and we compared the $^{131}$Xe(p,n), $^{127}$I($\alpha, \gamma$), $^{132}$Ba(p,x) $^{nat}$Ba(p,x), $^{133}$Cs(p,3n), $^{129}$Xe($\alpha$,2n), $^{nat}$Xe($\alpha$,xn) reactions. The yields of the most practical low and medium energy charged particle production routes \citep{TF2009b}, completed with our new experimental results on $^{133}$Cs(d,4n) reaction  are summarized in Fig. 11.
According to Fig. 11 out of the possible charged particle production routes the indirect production through the$^{132}$Ba(p,x) reaction is the most productive, but the isotopic abundance of the $^{132}$Ba is only 0.101 \%. The $^{133}$Cs(p,3n)$^{131}$Ba reaction also has high yield  and allows using targets with natural (mono)isotopic composition. The situation is similar in case of relying on the $^{133}$Cs(d,4n) reaction, but a  higher energy accelerator is required. The direct production of $^{131}$Cs through the $^{131}$Xe(p,n) reaction has also reasonable yield and this production route starts at the lowest energy.

\begin{figure}
\includegraphics[width=0.5\textwidth]{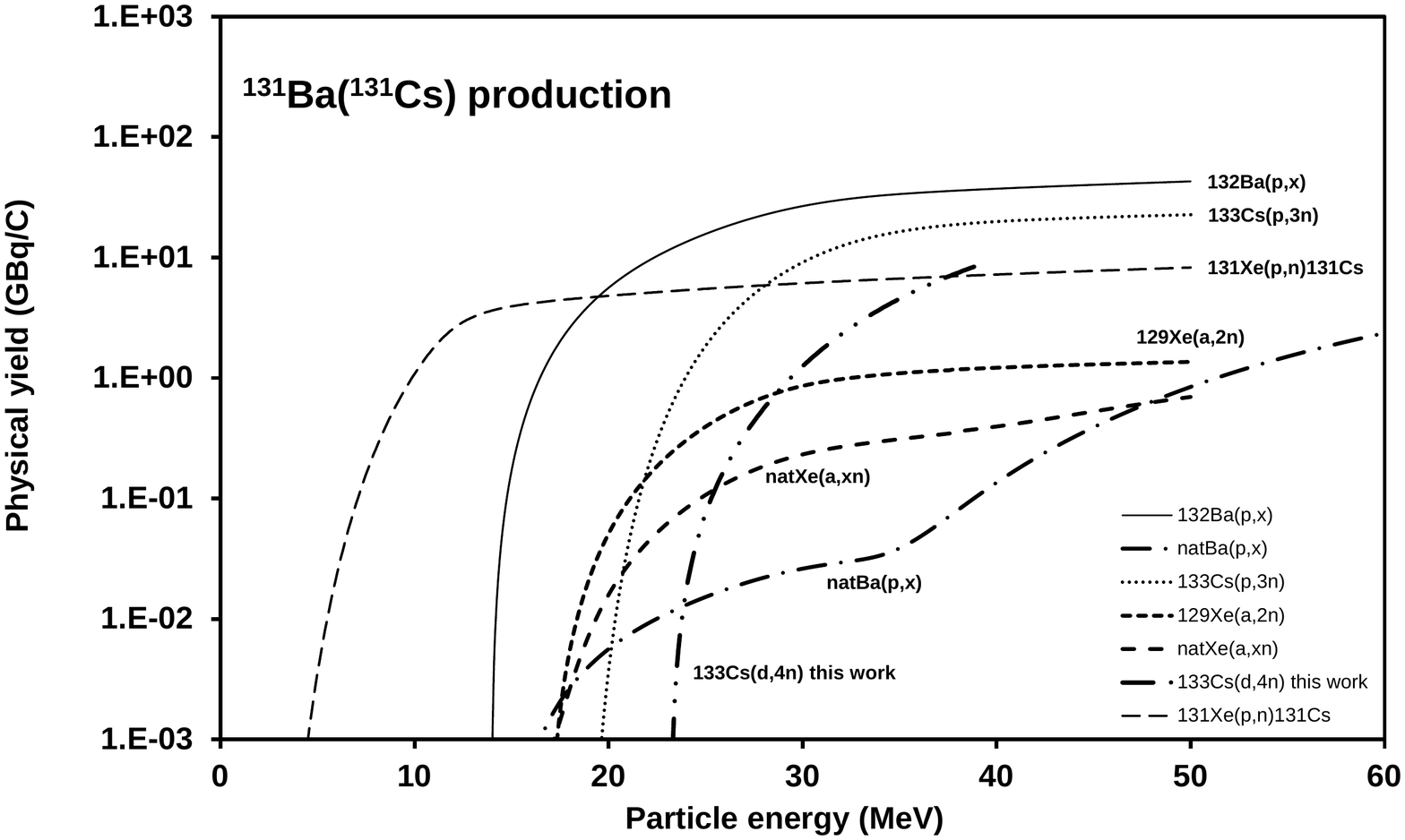}
\caption{Integral yields for production of $^{131}$Cs in $^{131}$Xe(p,n)  and $^{131}$Ba in $^{132}$Ba(p,x) $^{nat}$Ba(p,x), $^{129}$Xe($\alpha$,2n), $^{nat}$Xe($\alpha$,xn) and $^{133}$Cs(p,3n) and $^{133}$Cs(d,4n) (present work) reactions}
\label{fig:11}       
\end{figure}

\subsection{$^{133}$Ba}
\label{6.2}
The production of $^{133}$Ba via a (n,$\gamma$) reaction on $^{132}$Ba (which has a 0.101 \% natural abundance) in a nuclear reactor has been reported, but the product is not carrier free.
The cyclotron production of no carrier added $^{133}$Ba is possible using charged particle induced reactions: via (p,n) or (d,2n) reactions on monoisotopic $^{133}$Cs, via $^{132}$Xe($\alpha$,3n)  $^{nat}$Xe($\alpha$,xn), $^{131}$Xe($^3$He,n) and $^{nat}$Xe($^3$He,xn) reactions on gas targets using less frequently available and nowadays very expensive $^3$He beams.
The yields of the most important charged particle production routes, completed with our new experimental results on $^{133}$Cs(d,2n) reaction are summarized in Fig. 12. The 3He induced reactions are not presented in Fig. 12 because of the reason mentioned above. The yield calculations are based on fitted experimental cross section data, or TENDL-2014 theoretical results when experimental data are missing.
According to Fig. 12, out of the possible charged particle production routes, the $^{133}$Cs(p,n) and the $^{133}$Cs(d,2n) reactions are the most productive. In this range of mass numbers the maximum cross sections of the (p,n) and the (d,2n) reactions are comparable. At higher mass numbers the (d,2n) cross sections become significantly higher. 

\begin{figure}
\includegraphics[width=0.5\textwidth]{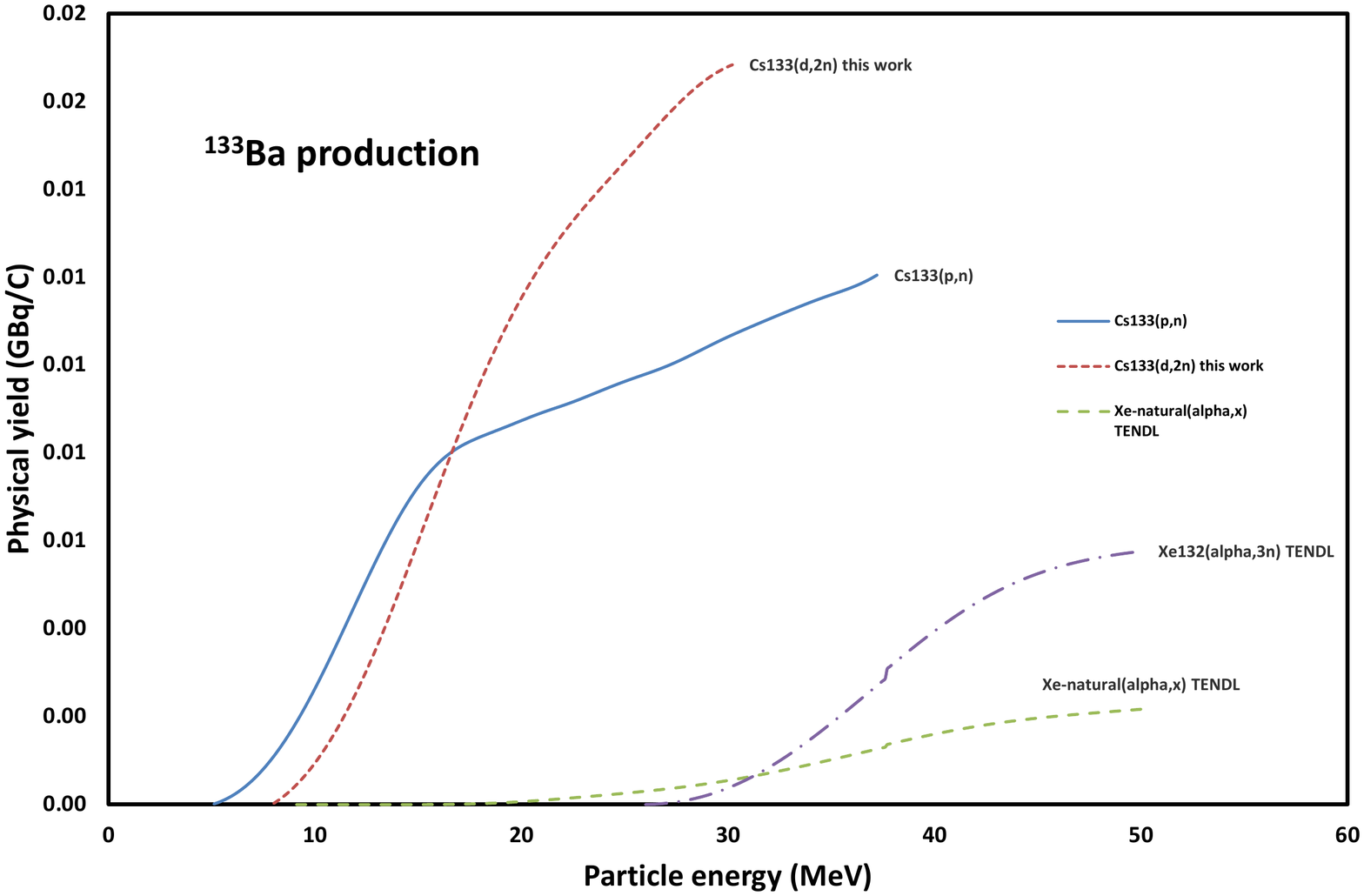}
\caption{Integral yields for production $^{132}$Xe($\alpha$,3n)  $^{nat}$Xe($\alpha$,xn), $^{133}$Cs(p,n) and $^{133}$Cs(d,2n)  (present work) reactions}
\label{fig:12}       
\end{figure}

\section{Summary and conclusions}
\label{7}
We present experimental cross sections for the nuclear reactions $^{133}$Cs(d,x)$^{133m,133mg,131mg}$Ba, $^{134,132}$Cs,  and $^{129m}$Xe up to 40 MeV deuteron  energies. The new data are first experimental activation data sets, except for $^{133}$Ba and $^{134}$Cs. The comparison with the recently published TENDL-2014 library shows only moderate agreement in few cases, and completely missing in other cases. Also the D-versions of the ALICE-IPPE and EMPIRE-II codes (specially adapted for deuteron induced reaction) are still largely deviant from the experimental data. 
The possible use of our experimental data for production $^{131}$Ba/$^{131}$Cs and $^{133}$Ba (section 6.) was discussed in comparison with other production routes, showing the importance of deuteron induced reactions in production of the above mentioned medically important radioisotopes.

\section{Acknowledgements}
\label{}
This work was done in collaboration of the Tohoku University (Sendai) and Institute for Nuclear Research (Debrecen). The authors acknowledge the support of the respective institutions in providing the beam time and experimental facilities.
 



\bibliographystyle{elsarticle-harv}
\bibliography{Csd}

\begin{thebibliography}{36}
\expandafter\ifx\csname natexlab\endcsname\relax\def\natexlab#1{#1}\fi
\expandafter\ifx\csname url\endcsname\relax
  \def\url#1{\texttt{#1}}\fi
\expandafter\ifx\csname urlprefix\endcsname\relax\def\urlprefix{URL }\fi

\bibitem[{Andersen and Ziegler(1977)}]{Andersen}
Andersen, H.~H., Ziegler, J.~F., 1977. Hydrogen stopping powers and ranges in
  all elements. The stopping and ranges of ions in matter, Volume 3. The
  Stopping and ranges of ions in matter. Pergamon Press, New York.

\bibitem[{Baron and Cohen(1963)}]{Baron}
Baron, N., Cohen, B.~L., 1963. Activation cross-section survey of
  deuteron-induced reactions. Physical Review 129~(6), 2636--2642.

\bibitem[{Belgya et~al.(2005)Belgya, Bersillon, Capote, Fukahori, Zhigang,
  Goriely, Herman, Ignatyuk, Kailas, Koning, Oblozinsky, Plujko, and
  Young}]{Belgya}
Belgya, T., Bersillon, O., Capote, R., Fukahori, T., Zhigang, G., Goriely, S.,
  Herman, M., Ignatyuk, A.~V., Kailas, S., Koning, A., Oblozinsky, P., Plujko,
  V., Young, P., 2005. Handbook for calculations of nuclear reaction data:
  Reference Input Parameter Library. http://www-nds.iaea.org/RIPL-2/. IAEA,
  Vienna.

\bibitem[{Bonardi(1987)}]{Bonardi}
Bonardi, M., 1987. The contribution to nuclear data for biomedical radioisotope
  production from the milan cyclotron facility.

\bibitem[{Dityuk et~al.(1998)Dityuk, Konobeyev, Lunev, and Shubin}]{Dityuk}
Dityuk, A.~I., Konobeyev, A.~Y., Lunev, V.~P., Shubin, Y.~N., 1998. New version
  of the advanced computer code alice-ippe. Tech. rep., IAEA.

\bibitem[{Dmitriev et~al.(1983)Dmitriev, Krasnov, and Molin}]{Dmitriev83}
Dmitriev, P.~P., Krasnov, N.~N., Molin, G.~A., 1983. Yields of radioactive
  nuclides formed by bombardment of a thick target with 22-mev deuterons. Tech.
  rep., INDC(CCP)-210/L.

\bibitem[{Dmitriev et~al.(1973)Dmitriev, Molin, and Panarin}]{Dmitriev73}
Dmitriev, P.~P., Molin, G.~A., Panarin, M.~V., 1973. Yields of ba133m and ba133
  and isomeric ratios for cs133(p, n)ba133m, g and cs133(d, 2n)ba133m,g. At.
  Energ. 35, 61.

\bibitem[{Eybel et~al.(1979)Eybel, Turner, Rayudu, Fordham, Friedman, and
  Messer}]{Eybel}
Eybel, C.~E., Turner, D.~A., Rayudu, G.~V., Fordham, E.~W., Friedman, A.~M.,
  Messer, J.~V., 1979. Myocardial imaging with 134mcesium. comparison of
  scintigraphic and cineradiographic results. Clin Cardiol 2~(3), 197--202.

\bibitem[{Henschke and Lawrence(1965)}]{Henschke}
Henschke, U.~K., Lawrence, D.~C., 1965. Cesium-131 seeds for permanent
  implants. Radiology 85, 1117.

\bibitem[{Herman et~al.(2007)Herman, Capote, Carlson, Oblozinsky, Sin, Trkov,
  Wienke, and Zerkin}]{Herman}
Herman, M., Capote, R., Carlson, B.~V., Oblozinsky, P., Sin, M., Trkov, A.,
  Wienke, H., Zerkin, V., 2007. Empire: Nuclear reaction model code system for
  data evaluation. Nuclear Data Sheets 108~(12), 2655--2715.

\bibitem[{Hermanne et~al.(2008)Hermanne, T\'ark\'anyi, and Tak\'acs}]{Hermanne}
Hermanne, A., T\'ark\'anyi, F., Tak\'acs, S., 2008. Production of medically
  relevant radionuclides with medium energy deuterons.

\bibitem[{Ignatyuk(2011)}]{Ignatyuk}
Ignatyuk, A.~V., 2011. Phenomenological systematics of the (d,p) cross
  sections,
  http://www-nds.iaea.org/fendl3/000pages/rcm3/slides//ignatyuk\_fendl-3

\bibitem[{Kinsey et~al.(1997)Kinsey, Dunford, Tuli, and Burrows}]{Kinsey}
Kinsey, R.~R., Dunford, C.~L., Tuli, J.~K., Burrows, T.~W., 1997. in Capture
  Gamma-Ray Spectroscopy and Related Topics, Vol. 2. (NUDAT 2.6
  http://www.nndc.bnl.gov/nudat2/). Vol.~2. Springer Hungarica Ltd, Budapest.

\bibitem[{Koning and Rochman(2012)}]{Koning2012}
Koning, A.~J., Rochman, D., 2012. Modern nuclear data evaluation with the talys
  code system. Nuclear Data Sheets 113, 2841.

\bibitem[{Koning et~al.(2014)Koning, Rochman, van~der Marck, Kopecky, Sublet,
  Pomp, Sjostrand, Forrest, Bauge, Henriksson, Cabellos, Goriely, Leppanen,
  Leeb, Plompen, and Mills}]{Koning2014}
Koning, A.~J., Rochman, D., van~der Marck, S., Kopecky, J., Sublet, J.~C.,
  Pomp, S., Sjostrand, H., Forrest, R., Bauge, E., Henriksson, H., Cabellos,
  O., Goriely, S., Leppanen, J., Leeb, H., Plompen, A., Mills, R., 2014.
  Tendl-2014: Talys-based evaluated nuclear data library.

\bibitem[{Mocoroa et~al.(1966)Mocoroa, Vignau, Caracoche, and
  Nassiff}]{Mocoroa}
Mocoroa, A., Vignau, H., Caracoche, M.~C., Nassiff, S.~J., 1966. Isomeric
  cross-section ratios for $^{134m}$cs, $^{134}$cs and $^{133m}$ba, $^{133}$ba
  formed in (d, p) and (d, 2n) reactions. J. Inorg. Nucl. Chem. 28, 5.

\bibitem[{Mukhammedov et~al.(1984)Mukhammedov, Vasidov, and
  Pardaev}]{Mukhammedov}
Mukhammedov, S., Vasidov, A., Pardaev, E., 1984. Use of proton and deuteron
  activation methods of analysis in the determination of elements with z $\>$
  42. Soviet Atomic Energy 56~(1), 56--58, tp379 Times Cited:4 Cited References
  Count:19.

\bibitem[{Murphy et~al.(2004)Murphy, Piper, Greenwood, Mitch, Lamperti,
  Seltzer, Bales, and Phillips}]{Murphy}
Murphy, M.~K., Piper, R.~K., Greenwood, L.~R., Mitch, M.~G., Lamperti, P.~J.,
  Seltzer, S.~M., Bales, M.~J., Phillips, M.~H., 2004. Evaluation of the new
  cesium-131 seed for use in low-energy x-ray brachytherapy. Medical Physics
  31~(6), 1529--1538.

\bibitem[{Natowitz and Wolke(1967)}]{Natowitz}
Natowitz, J.~B., Wolke, R.~L., 1967. Isomeric cross sections and yield ratios
  of (d,p) reactions below 15 mev. Phys. Rev. 155, 1352.

\bibitem[{Neirincky(1977)}]{Neirincky}
Neirincky, R.~D., 1977. Production of ba-133m for medical purposes.
  International Journal of Applied Radiation and Isotopes 28~(3), 323--325.

\bibitem[{of-Weights-and Measures(1993)}]{Error}
of-Weights-and Measures, I.-B., 1993. Guide to the expression of uncertainty in
  measurement, 1st Edition. International Organization for Standardization,
  Genève, Switzerland.

\bibitem[{Otuka and Tak\'acs(2015)}]{Otuka}
Otuka, N., Tak\'acs, S., 2015. Definitions of radioisotope thick target yields.
  Radiochimica Acta 103~(1), 1--6.

\bibitem[{Pement and Wolke(1966)}]{Pement}
Pement, F.~W., Wolke, R.~L., 1966. Compound-statistical features of
  deuteron-induced reactions (ii). the compound nucleus and
  stripping-evaporation mechanisms in (d,2n) reactions. Nucl. Phys. 86, 429.

\bibitem[{Pritychenko and Sonzogni(2003)}]{Pritychenko}
Pritychenko, B., Sonzogni, A., 2003. Q-value calculator.

\bibitem[{Sonzogni(2005)}]{Sonzogni}
Sonzogni, A.~A., 2005. Nudat 2.0: Nuclear structure and decay data on the
  internet, http://www.nndc.bnl.gov/nudat2. In: Haight, R.~C., Chadwick, M.~B.,
  Kawano, T., Talou, P. (Eds.), ND2004 - International Conference on Nuclear
  Data for Science and Technology. Vol. 769. AIP Conference Proceedings, pp.
  547--577.

\bibitem[{T\'ark\'anyi et~al.(2010)T\'ark\'anyi, Ditr\'oi, Kir\'aly, Tak\'acs,
  Hermanne, Yamazaki, Baba, Mohammadi, and Ignatyuk}]{TF2010}
T\'ark\'anyi, F., Ditr\'oi, F., Kir\'aly, B., Tak\'acs, S., Hermanne, A.,
  Yamazaki, H., Baba, M., Mohammadi, A., Ignatyuk, A.~V., 2010. Study of
  activation cross sections of proton induced reactions on barium: Production
  of ba-131 -> cs-131. Applied Radiation and Isotopes 68~(10), 1869--1877.

\bibitem[{T\'ark\'anyi et~al.(2011)T\'ark\'anyi, Hermanne, Ditr\'oi, Tak\'acs,
  Kir\'aly, Csikai, Baba, Yamazaki, Uddin, Ignatyuk, and Qaim}]{TF2011}
T\'ark\'anyi, F., Hermanne, A., Ditr\'oi, F., Tak\'acs, S., Kir\'aly, B.,
  Csikai, G., Baba, M., Yamazaki, H., Uddin, M.~S., Ignatyuk, A.~V., Qaim,
  S.~M., 25-28 Oct., 2010 2011. Systematic study of activation cross-sections
  of deuteron induced reactions used in accelerator applications.

\bibitem[{T\'ark\'anyi et~al.(2007)T\'ark\'anyi, Hermanne, Ditr\'oi, Tak\'acs,
  Szelecs\'enyi, Kir\'aly, Csikai, Sonck, Uddin, Hagiwara, Baba, Ido, Ohtshuki,
  Shubin, Kovalev, Dityuk, and Ignatyuk}]{TF2007}
T\'ark\'anyi, F., Hermanne, A., Ditr\'oi, F., Tak\'acs, S., Szelecs\'enyi, F.,
  Kir\'aly, B., Csikai, G., Sonck, M., Uddin, M.~S., Hagiwara, M., Baba, M.,
  Ido, T., Ohtshuki, T., Shubin, Y.~N., Kovalev, S.~F., Dityuk, A.~I.,
  Ignatyuk, A.~V., 2007. Cross sections of deuteron induced nuclear reactions
  on metal targets. In: Bersillon, O., Gunsing, F., Bange, E. e.~a. (Eds.),
  International Conference on Nuclear Data for Science and Technology. EDP
  Sciences, pp. 1027--1030.

\bibitem[{T\'ark\'anyi et~al.(2009{\natexlab{a}})T\'ark\'anyi, Hermanne,
  Kir\'aly, Tak\'acs, Ditr\'oi, Sonck, Kovalev, and Ignatyuk}]{TF2009}
T\'ark\'anyi, F., Hermanne, A., Kir\'aly, B., Tak\'acs, S., Ditr\'oi, F.,
  Sonck, M., Kovalev, S.~F., Ignatyuk, A.~V., 2009{\natexlab{a}}. Investigation
  of excitation functions of alpha induced reactions on natxe: Production of
  the therapeutic radioisotope $^{131}$cs. Nuclear Instruments \& Methods in
  Physics Research Section B-Beam Interactions with Materials and Atoms
  267~(5), 742--754.

\bibitem[{T\'ark\'anyi et~al.(2008{\natexlab{a}})T\'ark\'anyi, Hermanne,
  Tak\'acs, Ditr\'oi, Kir\'aly, Baba, Yamazaki, Kovalev, and Ignatyuk}]{TF2008}
T\'ark\'anyi, F., Hermanne, A., Tak\'acs, S., Ditr\'oi, F., Kir\'aly, B., Baba,
  M., Yamazaki, H., Kovalev, F.~S., Ignatyuk, A.~V., 2008{\natexlab{a}}.
  Investigation of production of the therapeutic radioisotope $^{131}$cs via
  charged particle induced reactions. In: 6th International Conference on
  Isotopes. p. 118.

\bibitem[{T\'ark\'anyi et~al.(2009{\natexlab{b}})T\'ark\'anyi, Hermanne,
  Tak\'acs, Rebeles, Van~den Winkel, Kir\'aly, Ditr\'oi, and
  Ignatyuk}]{TF2009b}
T\'ark\'anyi, F., Hermanne, A., Tak\'acs, S., Rebeles, R.~A., Van~den Winkel,
  P., Kir\'aly, B., Ditr\'oi, F., Ignatyuk, A.~V., 2009{\natexlab{b}}. Cross
  section measurements of the $^{131}$xe(p,n) reaction for production of the
  therapeutic radionuclide 131cs. Applied Radiation and Isotopes 67~(10),
  1751--1757.

\bibitem[{T\'ark\'anyi et~al.(1991)T\'ark\'anyi, Szelecs\'enyi, and
  Tak\'acs}]{TF1991}
T\'ark\'anyi, F., Szelecs\'enyi, F., Tak\'acs, S., 1991. Determination of
  effective bombarding energies and fluxes using improved stacked-foil
  technique. Acta Radiologica, Supplementum 376, 72.

\bibitem[{T\'ark\'anyi et~al.(2008{\natexlab{b}})T\'ark\'anyi, Tak\'acs,
  Ditr\'oi, Kir\'aly, Szelecs\'enyi, Hermanne, Van~den Winkel, Adam~Rebeles,
  Hilgers, Spahn, Qaim, Scholten, Uddin, Baba, Yamazaki, Ignatyuk, and
  Kovalev}]{TF2008b}
T\'ark\'anyi, F., Tak\'acs, S., Ditr\'oi, F., Kir\'aly, B., Szelecs\'enyi, F.,
  Hermanne, A., Van~den Winkel, P., Adam~Rebeles, R., Hilgers, K., Spahn, I.,
  Qaim, S.~M., Scholten, B., Uddin, M.~S., Baba, M., Yamazaki, H., Ignatyuk,
  A.~V., Kovalev, S.~F., 2008{\natexlab{b}}. Experimental study of the
  production routes of therapeutic radioisotopes at accelerators. In: 7th
  International Conference on Nuclear and Radiochemistry. NRC 7. p. 155.

\bibitem[{T\'ark\'anyi et~al.(2001)T\'ark\'anyi, Tak\'acs, Gul, Hermanne,
  Mustafa, Nortier, Oblozinsky, Qaim, Scholten, Shubin, and Youxiang}]{TF2001}
T\'ark\'anyi, F., Tak\'acs, S., Gul, K., Hermanne, A., Mustafa, M.~G., Nortier,
  M., Oblozinsky, P., Qaim, S.~M., Scholten, B., Shubin, Y.~N., Youxiang, Z.,
  2001. Beam monitor reactions (chapter 4). charged particle cross-section
  database for medical radioisotope production: diagnostic radioisotopes and
  monitor reactions. Tech. rep., IAEA.

\bibitem[{van~der Meulen et~al.(2010)van~der Meulen, Steyn, van~der Walt,
  Szelecs\'enyi, Kov\'acs, and Raubenheimer}]{Meulen}
van~der Meulen, N.~P., Steyn, G.~F., van~der Walt, T.~N., Szelecs\'enyi, F.,
  Kov\'acs, Z., Raubenheimer, H.~G., 2010. The isolation of ba-133 produced by
  proton induced reactions on cs using cation exchange chromatography. Journal
  of Radioanalytical and Nuclear Chemistry 285~(3), 491--498.

\bibitem[{van~der Walt et~al.(2008)van~der Walt, van~der Muelen, Steyn,
  Szelecs\'enyi, Kov\'acs, and Raubenheimer}]{Walt}
van~der Walt, T.~N., van~der Muelen, N.~P., Steyn, G.~F., Szelecs\'enyi, F.,
  Kov\'acs, Z., Raubenheimer, H.~G., 2008. The production of $^{133}$ba by
  proton induced reaction on cs. In: 6th international Conference on Isotopes.
  p. 236.

\end{thebibliography}







\end{document}